\documentstyle[psfig,conf_iap,]{article}
\def\lc{ $\lambda_c$ }
\def\bee{\begin{equation}}
\def\eee{\end{equation}}
\begin{document}
\heading{%
%
Did the dark energy  always have
a large Compton\\ wavelength?\\
%
} 
\par\medskip\noindent
\author{%
Bruce A. Bassett$^{1}$, Martin Kunz $^{2}$, David Parkinson$^{1}$ and 
Carlo Ungarelli$^{1,3}$
}
\address{%
Institute of Cosmology and Gravitation, University of 
Portsmouth, Portsmouth~PO1~2EG}
\address{%
Astronomy Centre, CPES, University of Sussex, Brighton, BN1 9QJ, UK}
\address{%
School of Physics and Astronomy, University of Birmingham
Edgbaston, Birmingham, B15 2TT, UK
}

\begin{abstract}
Until now all well-studied dark energy models proposed have very large Compton 
wavelengths, \lc $> 100 Mpc$ at all time and hence show no clustering 
on small scales.  Evidence against this would disfavour both $\Lambda$CDM and 
quintessence.  We show however, that large \lc is actually slightly 
{\em favoured}  by current CMB, LSS and SN1a data by comparing 
quintessence with a condensate scenario with the same 
background dynamics. The condensate scenario has a very small  
\lc before making a transition to 
negative pressure and large $\lambda_c$ and therefore exhibits
a rich  CMB phenomenology which roughly interpolates between
$\Lambda$CDM and flat CDM models. 
\end{abstract}
\section{Introduction}
In quintessence models the current acceleration of the universe is 
due to potential-dominated evolution of a scalar field today. This typically 
implies that the potential is very flat today and hence that the 
Compton wavelength, $\lambda_c = (V'')^{-1/2}$, is very large. This makes 
standard quintessence models difficult to distinguish from the simple 
cosmological constant ($\Lambda$) \cite{CBUC} since clustering will only 
occur on  scales larger than $\lambda_c$. Moreover,  
quintessence models fulfill the condition of a large \lc at all times. 
Therefore two interesting questions are:
\\

{\bf (1)}  {\em Can we build models which 
are not associated with a large \lc

either today or at earlier epochs? } 
\\

{\bf (2)} {\em Are such models favoured or disfavoured by cosmic data?}
\\ 

\noindent  
A strong affirmative to the second question would provide compelling evidence
against both $\Lambda$CDM and quintessence. In answer to the first 
question we propose the  idea of {\em condensate  dark energy} \footnote{Note
that k-essence for example, can lead to a strong variation in the speed of sound, potentially
leaving a detectable signal in the CMB \cite{sound}.}. 
In this scenario, part of 
the dark matter (which we take to be CDM) undergoes a transition to a 
condensate at some low redshift $z_t$, described by a scalar field with 
equation of state given by eq. (\ref{w}). Such condensate models 
have previously been envisaged \cite{cond1}, though with slightly 
different properties. 

To isolate the effects of the Compton wavelength we must be sure 
that our models have the same background 
dymanics. We therefore use the model-independent parametrisation 
of the equation of state of \cite{BKSU} (and generalised in \cite{CC}):
\bee
w(z) = \frac{w_f}{1+\exp((z-z_{t})/\Delta)}
\label{w}
\eee
where $w_f$ is the final value of the equation of state, $z_t$ is the redshift
at which $w(z)$ changes from $0$ to $w_f$ and $\Delta$ controls the rapidity 
of the transition. 

\begin{figure}
\centerline{\vbox{
\psfig{figure=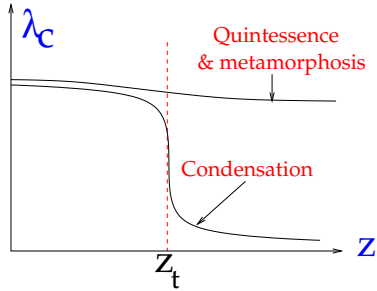,height=3.8cm}
}}
\caption[]{A schematic illustration of the evolution of the Compton wavelength 
$\lambda_c \equiv (V'')^{-1/2}$, as a function of redshift. In condensation
$\lambda_c$ jumps radically at the transition, $z_t$, while in metamorphosis
or quintessence models it is almost unchanged in cosmological terms. In 
our specific implementation $\lambda_c \sim 0$  for $z > z_t$ since we consider
a CDM condensate. 
}
\end{figure}
\begin{figure}
\centerline{\vbox{
\psfig{figure=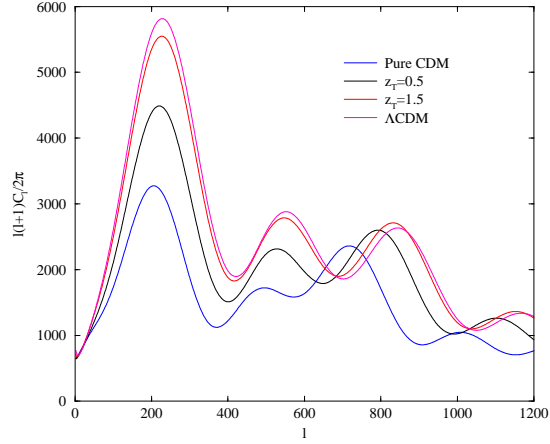,height=6.cm}
}}
\caption[]{The $C_{\ell}$ spectra for condensation as a function of
$z_t$.  In contrast to the case
of quintessence and metamorphosis the effect of $z_t$ is rather non-trivial for
$z_t < 1.5$, including an asymmetric offset pattern between the 2nd and 3rd 
peaks due to the induced change in $\Omega_b/\Omega_{cdm}$. For comparison 
pure CDM  is the lowest curve at $\ell = 200$, $\Lambda$CDM the highest. 
}
\label{cl}
\end{figure}
In the case of quintessence we can impose this equation of state to hold 
identically at all times. However, for condensation we require that for 
$z > z_t$ it hold only on average, i.e. $\langle w \rangle = 0$. This allows
the potential to be very curved and hence for \lc to be small and allow 
clustering on small scales.

\begin{figure}
\centerline{\vbox{
\psfig{figure=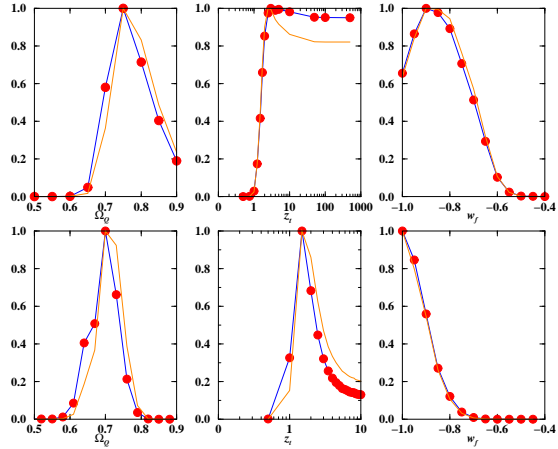,height=6.cm}
}}
\caption[]{The total 1-d likelihood plots for condensation (top) and
 metamorphosis (bottom row) for the three variables $\Omega_Q, z_t$ and $w_f$.
 These likelihood plots are computed from current CMB, LSS and SN1a data, all
 marginalised over  the Hubble constant (thick line with 
 red points). The thin solid 
 line corresponds to the case with a Gaussian prior $H = 72 \pm 8 km s^{-1} 
 Mpc^{-1}$. 
Condensation strongly disfavours $z_t < 1.5$, 
while metamorphosis favours $z_t \simeq 2$.
}
\label{1d}
\end{figure}

\section{Results}

To compare our models we use all recent CMB data sets (excluding Archeops),
the LSS power spectra inferred from 2df, PSCz, Abell/ACO and the binned SN1a 
data as described in \cite{BKSU} and \cite{cond}. 
We compute likelihoods over the 4d grid 
$(\Omega_Q, z_t, w_f, H$) where $H$ is the Hubble constant. We show the 
1d marginalised likelihoods for our $w(z)$ parameters in fig (\ref{1d}) for
both condensation and metamorphosis/quintessence. We also plot the same 
curves when we introduce a Gaussian prior for the Hubble constant 
coming from the  HST key project result: $H = 72 \pm 8$ km s$^{-1}$ Mpc$^{-1}$. 

Overall the condensate model has a better best-fit $\chi^2$ than 
the best $\Lambda$CDM model. Indeed, we found that for condensation 
the best-fit $\chi^2$ is  79.3 corresponding to 
$z_t=5\,,w_f=-0.95\,,\Omega_Q=0.75$, while for 
$\Lambda$CDM the best-fit $\chi^2$ is 84.9, 
(corresponding to $\Omega_{\Lambda}=0.73$). The best fits for metamorphosis
are even better with $\chi^2 = 78.8$ for $z_t = 1.5$ 
\cite{BKSU,cond}. 

To understand why condensation does not favour $z_t < 1.5$  we note that
this is due both to the CMB and LSS.  
Unlike quintessence models which have an almost trivial effect 
on the CMB (see e.g. \cite{BKSU}), varying $z_t$ in condensation affects 
the CDM density at decoupling and hence alters 
the gravitational potential wells. This effect
is only significant for $z_t < 2$ and boosts the height of the first 
acoustic peak relative to metamorphosis for $z_t = 0.5$ while giving an
asymmetric off-set pattern to the 2nd and 3rd peaks. 

The $C_{\ell}$ curves of fig. (\ref{cl}) can be understood by realising that
they interpolate between standard flat CDM (small $z_t$) 
and pure $\Lambda$CDM models (high $z_t$).  
While the total energy density of the universe is continuous across the 
transition surface at $z_t$, the CDM density is not, since some of it 
is rapidly converted into dark energy. 
Hence the ratio $\Omega_{b}/\Omega_{cdm}$ is {\em not} constant in this model,
but jumps across the transition at $z_t$. The smaller is $z_t$, the larger this
jump is, and hence the smaller is $\Omega_{b}/\Omega_{cdm}$, and {\em 
vice versa}. This leads to the characteristic pattern of asymmetric 2nd and 
3rd peaks seen for $z_t = 0.5$ due to the change in the gravitational potential
wells at recombination \cite{hu}. 

The LSS disfavours low $z_t$ since the turnover in the power spectrum 
occurs at too large a $k$ value, inconsistent with current observations 
\cite{cond}.


\section{Discussion}

In quintessence the Compton wavelength \lc of the scalar field is always 
very large, which implies that the transfer function of fluctuations is 
trivial on large scales. This is not necessary for a successful dark
energy model however and dropping this constraint yields a 
richer phenomenology.  
We consider one of the simplest models with a large change in \lc: {\em 
condensation}, and find that it is a worse fit to the data compared with
quintessence models if $z_t < 2$ due to both 
CMB and LSS constraints \cite{cond}. 

It is important to realise that our conclusions with regard to 
condensation would be altered if we assumed a different value of $\lambda_c$
initially by choosing our dark matter to be warm instead of cold. In that case
we would probably have a better fit to the data for $z_t < 2$, 
especially regarding LSS due to the well-documented problems of flat
CDM models in fitting both COBE and $\sigma_8$ constraints. 
Perhaps the most interesting result of condensation is that it opens up
the possibility of having dark energy which non-trivially alters the 
CMB and cluster-scale physics. Detailed analysis of condensation and 
comparison with quintessence/metamorphosis is
reported elsewhere \cite{cond}. 
\\

\vfill
\end{document}